 \documentstyle[12pt,titlepage]{article}
\begin{document}
\title{On dynamics of cylindrical and spherical relativistic domain walls
of finite thickness\thanks{Paper supported
in part by the grant KBN 2 P302 049 05.}}

 \vspace{0.5in}

\author{by\\
\\
H. Arod\'z   \\
Institute of Physics, Jagellonian University,
Cracow \thanks{Address: Reymonta 4, 30-059 Cracow, Poland.}
        \thanks{E-mail: UFARODZ@ztc386a.if.uj.edu.pl}\\
        \\
                and \\
                \\
         A.L. Larsen \\
        Nordita, Copenhagen \thanks{ Address: Blegdamsvej 17, DK-2100
        Copenhagen \O, Denmark.} \thanks{ E-mail: ALLARSEN@nbivax.nbi.dk}}

\date{September, 1993}

\vspace{3cm}
 \maketitle
\begin{abstract}
 Dynamics of cylindrical and spherical relativistic domain walls is
 investigated with the help of a new method based on Taylor expansion
 of the scalar field in a vicinity of the core of the wall. Internal
oscillatory
 modes for the domain walls are found. These modes are non-analytic in the
 "width" of the domain wall. Rather nontrivial transformation to a
 special coordinate system, widely used in investigations of relativistic
 domain walls, is studied in detail.

\end{abstract}

\section{Introduction}
During the last few years cosmic strings and domain walls have played an
important role in theoretical cosmology and astrophysics (for a review see
for instance references \cite{Vilenkin}, \cite{Kibble}). Most of the work has
been devoted to the cosmic strings since they are considered to be
cosmologically "safe". In contrary to this, it seems that stable domain
walls are cosmologically "disastrous". This was already pointed out by
Zeldovich et. al. \cite{Zeldovich} who considered domain wall structures
in models with spontaneous breaking of CP-symmetry. They argued that the
energy density of the domain walls is so large that they would dominate
the universe completely,
violating the observed approximate isotropy and homogenity. In other words,
the domain walls were assumed to somehow disappear again soon after their
creation in the early universe, for instance by collapse, evaporation or
simply by inflating away from our visible universe.

Much later however, Hill et. al. \cite{Hill} introduced the so-called
"light" or "soft" domain walls. They considered a late-time
(post-decoupling) phase transition and found that light (and thick)
domain walls could be produced, that were not necessarily in contradiction
with the observed large-scale structure of the universe.

Anyway, whatever the cosmological and astrophysical effects are, we find it
important to obtain a better understanding of the dynamics of domain walls
because of
the simple fact, that if the potential of a scalar field has at least
two separated degenerate minima the domain walls are there!
\vskip 6pt
\hspace*{-6mm}In this paper we will consider the dynamics of domain walls in
the simplest model of a scalar field with a $\Phi^4$ potential with two
degenerate minima. This problem can be (and has been) attacked in several
different ways. One possibility is to work directly with the field
equations of the scalar field in the general case, i.e. without making
any simplifying assumptions about the field. In that case the wall is nothing
but a special field configuration and because of the non-linearity of the
field equations one can usually not do anything analytically, and must
therefore rely on computer simulations. A completely different and quite
popular approach is to forget everything about the underlying field theory
and simply treat the domain wall as a minimal 3-surface (3-membrane)
embedded in 4-dimensional Minkowski space-time. The domain wall is then
described by a generalized Nambu-Goto action, and the corresponding
equations of motion determine the dynamics of the wall itself (in terms of
wall-coordinates).

During the last 4-5 years there has been several attemps to make a
"compromise" of the two approaches described above, in the sense that one
would like to have an approach where the wall itself is somehow described
by a hypersheet-like object moving around in spacetime, but where the
scalar field is also somehow present. The calculations of the dynamics
of the walls are then done analytically, but one has to make some
simplifying assumptions about the geometry of the walls (spherical,
cylindrical, plane, locally plane,$\ldots$) and about the scalar field in the
vicinity of the walls. Before
describing the present work let us review two of these previous
approaches.

Silveira \cite{Silveira1} considered the evolution of a spherically
symmetric domain wall by defining an average radius
\begin{eqnarray}
<r>\equiv\frac{\int\rho r dV}{\int\rho dV},\nonumber
\end{eqnarray}
where $\rho=\rho(\Phi)$ is the energy density of the scalar field. To actually
calculate $<r>$ Silveira took as a first approximation the well-known
\cite{Zeldovich} plane wall scalar field solution, and was then able to obtain
a formula relating $<\ddot{r}>$, $<\dot{r}>$ and $<r>$ (where a dot denotes
derivative with respect to time), thus generalizing the equation of motion
for a spherically symmetric Nambu-Goto wall \cite{Letelier}.

Gregory et. al. \cite{Gregory,Gregory2,Gregory3}
(see also \cite{Silveira,Larsen,Barrabes})
generalized a method used for cosmic strings
\cite{Maeda,Gregory4} to consider
the dynamics of locally plane domain walls. Making an
expansion in the thickness of the wall, they were able to separate the
field equations of the scalar field, and to translate the result into a
Nambu-Goto wall with an extra rigidity term
\begin{eqnarray}
S=\mu_0\int\sqrt{-g}(1+\alpha R)d^3\sigma,\nonumber
\end{eqnarray}
where $g$ is the determinant of the induced metric on the
world-hypersheet of the wall, $R$ is
the corresponding Ricci-curvature and $\alpha$ is a constant. To make the
separation of the field equations, one has to take the derivative in the
direction normal to the wall, to be at least one order lower in the expansion
parameter, than the derivatives in the tangential directions. This is
certainly fulfilled for the static plane wall \cite{Zeldovich}, but is
merely an assumption in the other cases. Furthermore, it is not really
clear whether the expansion in the thickness is meaningful at all, when it
comes to the time evolution of the walls. This point will be explained in
more detail in Section 2.
\vskip 6pt
\hspace*{-6mm}The purpose of the present paper is to develop yet another
approach to the dynamics of domain walls. For convenience we will only
consider cylindrical and spherical domain walls. The core of the domain wall
is in our approach defined by the location of zeros of the scalar field, so
we will look for solutions to the field equations, invoking the relevant
symmetries and appropriate boundary conditions. An equation determining
the dynamics of the core of the wall itself (in terms of wall coordinates)
will then come out as a byproduct.

Our method is explained in detail in Section 2 for the case of a cylindrical
wall. The case of a spherical wall proceeds similarly and is therefore briefly
presented in the Appendix. In Section 3 we consider the time evolution of the
cylindrical wall and especially the time evolution of the scalar
field in the vicinity of the core of the wall. In Section 4 we consider
 some of the results of Sections 2 and 3 when transformed to the laboratory
  frame. In Section 5 we close the paper with some remarks, in particular
  we point out merits and demerits of our approach.

\section{The equation of motion and the method of solving it}

We will consider one of the simplest field-theoretical models (in
four-dimen\-sional, flat Minkowski space-time) in which the
domain wall type solutions appear \cite{Vilenkin}.
The model involves a single, real-valued scalar field
$\Phi(x)$ with Lagrangian given by the following formula
\begin{equation}
L=-\frac{1}{2} \eta_{\mu\nu}\partial^{\mu}\Phi\partial^{\nu}\Phi - V(\Phi),
\end{equation}
where $(\eta_{\mu\nu})=\mbox{diag}(-1,1,1,1)$, and
\begin{equation}
V(\Phi)= \frac{\lambda}{2}(\Phi^{2}-v^{2})^{2},
\end{equation}
$\lambda$ and $v$ are positive constants. The field equation following from
this Lagrangian is
\begin{equation}
\partial_{\mu}\partial^{\mu}\Phi - \frac{\partial V}{\partial \Phi} = 0,
\end{equation}
and the conserved energy-momentum tensor has the form
\begin{equation}
T^{\mu\nu}=\partial^{\mu}\Phi \partial^{\nu}\Phi + \eta^{\mu\nu} L .
\end{equation}

The classical ground states are given by $\Phi=\pm v$. The domain walls
 arise if there are regions in the space where the field
$\Phi$ has different vacuum values - the domain walls are solutions of Eq.(3)
interpolating between such regions. The example is provided by the
planar domain wall solution of Eq.(3) \cite{Zeldovich}:

\begin{equation}
\Phi(x^{1},x^{2},x^{3},x^{0}) = v\; \tanh(\alpha x^{3}),
\end{equation}
where $\alpha = \sqrt{\lambda v^{2}}$. It describes the static wall located at
the $x^{3}=0$ plane. In this case the energy density $T^{00}$ is
 exponentially localised at
that plane. For $x^{3}\rightarrow \pm\infty $ the field $ \Phi(x)$ approaches
its vacuum values $\pm v$.
The wall has finite thickness of the order $2\alpha^{-1}$.
 In general the domain
walls are not static, and it is a very interesting problem to describe
their evolution.

In the present paper we will consider evolution of a
domain wall between a cylindrical region
around the $x^{3}$-axis in which $\Phi \cong -v$ and the remaining part of the
space where $\Phi \cong +v$. Also, in the Appendix, we briefly discuss the
much similar case of a spherical wall separating the region around the
origin $\vec{x}=0$ (in which $\Phi\cong -v$) from the remaining part of
the space (where $ \Phi \cong +v $). These domain walls are non-static.

The first, rather standard step in investigations
 of the evolution of the domain
walls is transformation of the field equation (3) to special coordinates
\cite{Gregory}-\cite{Gregory4} which
can be regarded as a generalisation of Fermi coordinates
used in the general relativity \cite{Synge}. The rather complicated relation
of these coordinates to the laboratory frame coordinates we describe in
detail in Section 4. The characteristic (in fact, the defining) feature
of the simple domain wall solutions we consider in this paper is location of
topological zeros of the field $\Phi$ on a  smooth surface in space
at each moment of time (in our case it is a cylinder or
a sphere). This surface we shall call the core of the domain wall.
{}From the space-time point of view the core sweeps a
3-dimensional submanifold  $\Sigma$ in Minkowski space-time, the
world-hypersheet of the domain wall. In the case of
cylindrical domain wall, it can be parametrised in the following manner
\begin{equation}
\left(
\begin{array}{c}
      X^{0} \\
      X^{1} \\
      X^{2} \\
      X^{3}
\end{array}
\right) (\tau,\sigma,\zeta) = \left( \begin{array}{c} \tau \\
r(\tau) \; \cos \sigma \\
r(\tau) \; \sin \sigma \\
\zeta
\end{array}
\right),
\end{equation}
where $\tau$ is the time coordinate for the points on the cylinder and
the parameters $\sigma,\zeta$ parametrise the cylinder: $0\leq\sigma<2\pi$ and
$-\infty<\zeta<+\infty$.

 The special
coordinates in the Minkowski space-time are $(\tau,\sigma,\zeta,\xi)$.
They are introduced by the formula
\begin{equation}
x^{\mu} = X^{\mu}(\tau,\sigma,\zeta) + \xi \; n^{\mu},
\end{equation}
where $x^{\mu}$ are the Cartesian, laboratory frame coordinates in the
space-time, and $n$ is a normalised, space-like four-vector orthogonal to
$\Sigma$, i.e.
\[ n^{2}=1, \;\;\; n_{\mu}X^{\mu}_{,a}=0, \]
where $a=\tau,\sigma,\zeta$. In the following we shall take
\[ (n^{\mu})=\frac{1}{\sqrt{1-\dot{r}^{2}}}(\dot{r}, \cos{\sigma},
\sin{\sigma}, 0). \]
 For the points on $\Sigma$ the coordinate
$\xi=0$. The Minkowski space-time metric tensor
 in the new coordinates is diagonal.
It has the following non-vanishing components
\begin{equation}
G_{\tau\tau}= - (\sqrt{1-\dot{r}^{2}}+\frac{\xi\ddot{r}}{1-\dot{r}^{2}})^{2},
\;\;
G_{\sigma\sigma}=(r+\frac{\xi}{\sqrt{1-\dot{r}^{2}}})^{2},
\end{equation}
\[    G_{\zeta\zeta}=G_{\xi\xi}=1,  \]
 where the dot denotes the derivative with respect to $\tau$. For the
 four-volume element we have the formula
  \begin{equation}
  d^{4}x = \sqrt{-G} \; d\tau d\sigma d\zeta d\xi,
  \end{equation}
  where
  \begin{equation}
  \sqrt{-G}=(\sqrt{1-\dot{r}^{2}}+\frac{\xi\ddot{r}}{1-\dot{r}^{2}})\;(r+
  \frac{\xi}{\sqrt{1-\dot{r}^{2}}}).
   \end{equation}
 We see  from this formula that the new coordinates are well-defined if
 $\dot{r}^{2} < 1$ and
 \begin{eqnarray}
 -r\sqrt{1-\dot{r}^{2}} < \xi, \;    & -(1-\dot{r}^{2})^{3/2} < \xi \ddot{r}.
 \end{eqnarray}
 The coordinate $\xi$ is related to the distance from the $x_{3}$-axis
 \begin{equation}
 \frac{\xi}{\sqrt{1-\dot{r}^{2}}}= \sqrt{x^{2}_{1}+x^{2}_{2}}-r(\tau),
 \end{equation}
 as it is easily seen from formulae (6),(7). The new coordinates are not
 valid in particular on the $x_{3}$-axis.
 The d'Alembertian $\partial_{\mu}\partial^{\mu}\Phi$, present on the
 l.h.s. of Eq.(3), is given in the new coordinates by the standard formula
 \[ \partial_{\mu}\partial^{\mu}\Phi =\frac{1}{\sqrt{-G}}\partial_{\tau}(
 \sqrt{-G}G^{\tau\tau}\partial_{\tau}\Phi) +
 G^{\sigma\sigma}\partial^{2}_{\sigma}\Phi
 +\partial^{2}_{\zeta}\Phi +\frac{1}{\sqrt{-G}}\partial_{\xi}(\sqrt{-G}
 \partial_{\xi}\Phi),  \]
 where $G^{\tau\tau},\;G^{\sigma\sigma}$ are equal to $(G_{\tau\tau})^{-1},
  \;(G_{\sigma\sigma})^{-1}$,  respectively.

 The formulae given above contain the unspecified function $r(\tau)$.
 It turns out that this function can essentially
 be determined from the requirement
 that the field $\Phi$ vanishes on $\Sigma$, i.e.
 \begin{equation}
 \Phi(\tau,\sigma,\zeta,\xi)\mid_{\xi =0}=0.
 \end{equation}

  The assumption of cylindrical symmetry means that the field $\Phi$ does
  not depend on the coordinates $\sigma$ and $\zeta$. Then, the field
  equation (3) simplifies to
  \begin{equation}
  \frac{1}{\sqrt{-G}}\partial_{\tau}(\sqrt{-G}G^{\tau\tau}\partial_{\tau}\Phi)
  +\frac{1}{\sqrt{-G}}\partial_{\xi}(\sqrt{-G}\partial_{\xi}\Phi)
  -2\lambda(\Phi^{2}-v^{2})\Phi=0.
  \end{equation}

  This equation is too complicated in order to hope for finding
  exact solutions. Of possible approximation schemes, in literature prevails
  expansion in the "width" of the domain wall $\alpha^{-1}$
   \cite{Gregory}--\cite{Barrabes}. The quotation mark is due to the
   fact that the actual width of the domain wall is in general
   time-dependent and not equal to $\alpha^{-1}$.
  One rescales the field $\Phi$ and the coordinate $\xi$,
  \[  \Phi= v\phi, \;\;\; \xi=s/\alpha,  \]
   and one writes Eq.(14)  in the form
  \begin{equation}
  \frac{1}{\alpha^{2}}\frac{1}{\sqrt{-G}}\partial_{\tau}(\sqrt{-G}G^{\tau\tau}
  \partial_{\tau}\phi) + \frac{1}{\sqrt{-G}}\partial_{s}(\sqrt{-G}\partial
  _{s}\phi) - 2(\phi^{2}-1)\phi=0.
  \end{equation}
 Next, one expands the solution $\phi$ in powers of $\alpha^{-1}$:
 \begin{equation}
 \phi = \phi^{(0)} + \alpha^{-1}\phi^{(1)} + \alpha^{-2} \phi^{(2)} + \ldots .
 \end{equation}
 The unpleasant feature of this approach is that the part of Eq.(15)
 containing the $\tau$-derivatives, and therefore responsible for the
 evolution of the domain wall, is regarded as a perturbation. In such
 cases the expansion of the type (16) is in general not satisfactory.
 The problem is known in mathematical literature under the name "singular
 perturbations" \cite{Simon,Hinch}. It can be illustrated with a
 simple example provided by the harmonic oscillator equation:
 \begin{equation}
 \alpha^{-2}\ddot{\phi}(\tau)=-\phi(\tau).
 \end{equation}
It is easy to see that the expansion (16) gives
\begin{equation}
\phi^{(0)}=\phi^{(1)}=\phi^{(2)}=...=0,
\end{equation}
while the general solution of Eq.(17), $\phi\sim\cos{\alpha(\tau-\tau_{0})}$,
is not analytic in $\alpha^{-1}$ for $\alpha^{-1}\rightarrow 0$. However,
notice that $\phi =0$ obtained with the help of the expansion (16) is a
particular solution of Eq.(17). To the best of our knowledge, the
problem whether the expansion (16) does approximate some solution
of the Eq.(3) or not, has not been discussed in literature.

On the basis of knowledge about wave equations one can expect that Eq.(3)
possesses solutions with a component of the oscillatory type with the
characteristic frequency proportional to $\alpha$. Such solutions are certainly
missed by the expansion in powers of $\alpha^{-1}$. On the other hand,
because of its nonlinearity, Eq.(3) possesses nontrivial static solutions,
e.g. the planar domain wall (5), without the oscillatory type component.
The static solution can be regarded as compatible with the expansion
in positive powers of $\alpha^{-1}$ -- in this case $\phi^{(i)}=0$ for $i>0$.
Now, the question is whether the cylindrical domain wall solutions contain the
oscillatory component. If not, the solutions can probably be obtained
in the form of the series (16). To find out the answer one has to
investigate the evolution of the cylindrical domain wall using a method which
is not insensitive to the presence of the oscillations.

The method we propose in this paper is based on Taylor expansion of
$\Phi(\tau,\xi)$ in the variable $\xi$. Analogous approach has been used
in \cite{Arodz} to investigate dynamics of vortex lines. We assume
that the domain wall
solution is smooth, therefore it is natural to expect that the solution can be
approximated by Taylor formula. If we consider a finite, sufficiently small
interval $[-\xi_{1},\;\xi_{0}]$ ($\xi_{1},\xi_{0}$ are positive) of the
$\xi$ variable, we may hope that
the first few terms of the Taylor formula would give a good approximation.
The advantage such polynomial approximation offers is calculational
simplicity. The coefficients of the expansion can depend on $\tau$.
 Thus, in the interval $[-\xi_{1},\;\xi_{0}]$
\begin{equation}
\Phi(\tau,\xi)=a(\tau)\xi +\frac{b(\tau)}{2}\xi^{2}+\frac{c(\tau)}{3!}\xi^{3}
+...
\end{equation}
Formula (19) secures vanishing of $\Phi$ for $\xi=0$, condition (13).

The domain wall solution is characterised by the fact that for sufficiently
large $|\xi|$ the field $\Phi$ approaches its vacuum values $\pm v$. In order
to find the domain wall solution in the whole range of the $\xi$ variable,
one should patch at $\xi=-\xi_{1},\xi_{0}$
the solution (19) with asymptotic solutions for large $|\xi|$.
In order to avoid overloading our paper with formulae, we choose the simplest
possibility: we assume that
\begin{equation}
\begin{array}{c}
\Phi(\tau,\xi)=-v \;\;\mbox{for}\;\; \xi\leq-\xi_{1}, \\
\Phi(\tau,\xi)=+v \;\;\mbox{for}\;\; \xi\geq\xi_{0}.
\end{array}
\end{equation}
It is quite straightforward to extend our considerations by replacing
 (20) by more accurate asymptotic solutions with exponential
 $(\exp{(-2\alpha\xi)})$ corrections to the vacuum values $\pm v$.

 The patching conditions at $\xi=- \xi_{1},\xi_{0}$ follow in a
 standard manner from Eq.(14): one integrates Eq.(14) over $\xi$
 in arbitrarily small intervals $[-\xi_{1}-\epsilon,\;-\xi_{1}+\epsilon]$,
 $[\xi_{0}-\epsilon,\;\xi_{0}+\epsilon]$ and one lets
  $\epsilon \rightarrow 0$. With the asymptotic form (20) taken
  into account, they have the following form
  \begin{equation}
  \begin{array}{cc}
  \Phi(\tau,\xi_{0})=v, & \partial_{\xi}\Phi\mid_{\xi=\xi_{0}}=0, \\
  \Phi(\tau,-\xi_{1})=-v, & \partial_{\xi}\Phi\mid_{\xi=-\xi_{1}}=0.
  \end{array}
  \end{equation}
  The functions $a(\tau),\;b(\tau),\;c(\tau)$ present in formula (19)
  are related to each other because of recurrence relations following
  from inserting the expansion (19) into Eq.(14) and equating to zero
  coefficients in front of successive powers of $\xi$. This gives
  the following relations
  \begin{equation}
  b=-a\frac{1}{\sqrt{1-\dot{r}^{2}}}\left(\frac{1}{r}
  +\frac{\ddot{r}}{1-\dot{r}^{2}}\right),
  \end{equation}
  \begin{eqnarray}
  c=\frac{1}{1-\dot{r}^{2}}\ddot{a} + \frac{\dot{a}\dot{r}}{1-
  \dot{r}^{2}}\left(\frac{1}{r}+\frac{\ddot{r}}{1-\dot{r}^{2}}\right)

+\frac{a}{1-\dot{r}^{2}}\left(\frac{1}{r}+\frac{\ddot{r}}{1-\dot{r}^{2}}\right)^{2} \\

+\frac{a}{1-\dot{r}^{2}}\left(\frac{1}{r^{2}}+\frac{\ddot{r}^{2}}{(1-\dot{r}^{2})^{2}}\right)
  -2\lambda v^{2} a,  \nonumber
  \end{eqnarray}
  and similar ones for other coefficients in expansion (19). However, in the
  present paper we will restrict our considerations to the first three terms
  in expansion (19). Similarly as for the asymptotic solution for large
  $|\xi|$, there is no difficulty in including the higher powers of
  $\xi$ in the expansion (19), and thus improving the accuracy of our
  solution.

  The patching conditions (21) now take the form
  \begin{equation}
  a\xi_{0} +\frac{b}{2}\xi^{2}_{0} +\frac{c}{3!}\xi^{3}_{0}=v,
  \end{equation}
  \begin{equation}
  a+b\xi_{0}+\frac{c}{2}\xi^{2}_{0}=0,
  \end{equation}
  \begin{equation}
  -a\xi_{1}+\frac{b}{2}\xi^{2}_{1}-\frac{c}{3!} \xi^{3}_{1}=-v,
  \end{equation}
  \begin{equation}
  a-b\xi_{1}+\frac{c}{2}\xi^{2}_{1}=0.
  \end{equation}
  It is easy to see that these conditions imply that
  \begin{equation}
  \xi_{0}=\xi_{1},\;\; a=\frac{3v}{2\xi_{0}},\;\; b=0,\;\;c=\frac{-3v}
  {\xi^3_0}=-\frac{8a^{3}}{9v^{2}}.
  \end{equation}
Comparing this with formula (22) we obtain an equation for the radius
 $r(\tau)$ of the cylindrical domain wall:
 \begin{equation}
 \frac{\ddot{r}}{1-\dot{r}^{2}} +\frac{1}{r}=0.
 \end{equation}
 It turns out that this equation is a particular case of Euler-Lagrange
 equation for a relativistic membrane, obtained from
 the Nambu-Goto action
 \[ S= \int d\tau d\sigma d\zeta \sqrt{-g}, \]
 where $g\equiv G\mid_{\xi=0}$ is the determinant of the induced metric
 $g_{ab}\equiv G_{ab}\mid_{\xi =0}$ on the world-hypersheet $\Sigma$.
 In order to obtain Eq.(29) from the Euler-Lagrange equation it is
 sufficient to assume that the membrane has the form of the cylinder.
 Thus, within our method of solving Eq.(14) and within the chosen
  approximations, the evolution of the core of the cylindrical domain wall
  is described by the Nambu-Goto equation (29). This result is in agreement
  with results of earlier investigations of dynamics of the domain walls
  in which other methods were employed, see e.g. \cite{Silveira1},
  \cite{Gregory3}.

  The $\tau$-dependence of the coefficient $a(\tau)$ follows from comparison
  of the relation (23) with the patching condition (27). Taking into account
  formulae (28) and Eq.(29), we obtain the following equation
  \begin{equation}
  \frac{\ddot{a}}{1-\dot{r}^{2}}+ \frac{2a}{r^{2}(1
  -\dot{r}^{2})}-2  \lambda v^{2}a + \frac{8}{9v^{2}}a^{3}=0,
  \end{equation}
  which we shall analyse in the next Section.

  The formulae (28) and the equations (29),(30) give the full set of
  implications of the patching conditions (24)--(27) and of the recurrence
  relations (22),(23). The function $\xi_{0}(\tau)$ can be regarded as the
  half-width of the domain wall in the $\xi$ coordinate. How this width
  transforms to the laboratory frame we discuss in Section 4. We see from
  (28),(30) that the width is $\tau$-dependent. We shall show in
  the next Section that in general it contains the oscillating
  component which is non-analytic in $\alpha^{-1}$ for
  $\alpha\rightarrow\infty$ and therefore can
  not be found in perturbative expansion in positive powers of $\alpha^{-1}$.
  However, we also find out that there is a possibility for "fine tuning"
  initial data for Eq.(30) in such a way that the oscillating component does
  not appear. This very particular solution can probably be obtained by the
  perturbative expansion in positive powers of $\alpha^{-1}$.

\section{The time-dependence of the width of the domain wall}

 The half-width $\xi_{0}(\tau)$ of the domain wall is given in our approach by
 formula
 \begin{equation}
 \xi_{0}(\tau)=\frac{3v}{2a(\tau)},
 \end{equation}
 where $a(\tau)$ obeys Eq.(30).

 Equation (30) involves the function $r(\tau)$ which should be determined
 from the Nambu-Goto equation (29). General solution of the Nambu-Goto
 equation is easy to obtain when the corresponding "energy" integral of
 motion is taken into account:
 \begin{equation}
 E=\frac{r}{\sqrt{1-\dot{r}^{2}}}.
 \end{equation}
 Then, Eq.(29) can be written in the form
 \begin{equation}
 \ddot{r}=-\frac{1}{E^{2}}r,
 \end{equation}
 and we see that its general solution  has the form
 \begin{equation}
 r(\tau)=r_{0}\cos{\frac{\tau -\tau_{0}}{E}},
 \end{equation}
 where, as follows from formulae (32),(34),
 \begin{equation}
 E=r_{0}.
 \end{equation}
 The "velocity" $\dot{r}(\tau)$ vanishes at $\tau=\tau_{0}$. In principle
$r_{0}$ is the amplitude of harmonic oscillations of
 the core of the wall, but actually our analysis of the motion of the domain
 wall breaks down for small $r$. The problem lies in the use of the special
 coordinates. They are well defined if the conditions (11) are satisfied.
 For $r(\tau)$ given by formula (34) these conditions can be written in the
 form
 \begin{equation}
 \frac{r^{2}}{r_{0}}>\xi>-\frac{r^{2}}{r_{0}}.
 \end{equation}
 Our analysis  ceases to be sensible if the allowed range of $\xi$
 is smaller than $\xi_{0}$. This gives the restriction
 \begin{equation}
 \frac{r^{2}(\tau)}{r_{0}}>\xi_{0}(\tau),
 \end{equation}
 which certainly excludes $r(\tau)=0$. Thus, we can follow the evolution of the
 domain wall only for a part of the first quarter of the cycle of the
 oscillations of the core predicted by the Nambu-Goto equation, i.e.
 for $\tau$ in the interval (for convenience we put $\tau_0=0$)
 \begin{equation}
 0\leq \tau < (\frac{\pi}{2}-\delta)r_{0},
 \end{equation}
 where $\delta$ is to be determined from (37) after $\xi_{0}(\tau)$ is found.

 Let us write Eq.(30) with formulae (34),(35) taken into account
 \begin{equation}
 \frac{\ddot{a}}{\cos^{2}{\frac{\tau}{r_{0}}}} +
 \frac{2a}{r_{0}^{2}\cos^{4}{\frac{\tau}{r_{0}}}} - 2\lambda v^{2}a
 +\frac{8}{9v^{2}}a^{3} =0.
 \end{equation}
 Next, we pass to a dimensionless function $A$,
 \begin{equation}
 a(\tau)\equiv \frac{3}{2} v \alpha A(t),\;\; \mbox{i.e.}\;\;
 A(t)=\frac{1}{\alpha\xi_{0}(\tau)},
 \end{equation}
 of a dimensionless variable $t$:
 \begin{equation}
 t\equiv \frac{\tau}{r_{0}}.
 \end{equation}
 Let us also introduce a dimensionless parameter $\kappa$,
 \begin{equation}
 \kappa\equiv r_{0}\alpha,
 \end{equation}
 where $\alpha\equiv\sqrt{\lambda v^{2}}$. $\kappa$ gives the
 initial radius of the core of the cylindrical domain wall in the $\alpha^{-1}$
  units, while $A$ gives the inverse of the half-width of the wall
 in these units.
 Then, Eq.(39) can be written in the following form
 \begin{equation}
 \frac{1}{2\kappa^{2}}\frac{1}{\cos^{2}{t}}\frac{d^{2}A}{dt^{2}}=- \frac{1}
 {\kappa^{2}}\frac{A}{\cos^{4}{t}} + A - A^{3}  .
 \end{equation}
 The interval of the $t$ variable we are interested in corresponds to the
 interval of the $\tau$ parameter given by (38), i.e.
 \begin{equation}
 0\leq t < \frac{\pi}{2}-\delta.
 \end{equation}

The equation (43) is of  anharmonic oscillator type, but it has
t-dependent coefficients. Because of finiteness of the interval of the t
variable, see (44), Eq.(43) is relatively simple for numerical analysis.
Examples of numerical solutions, obtained with the help of Mathematica,
are presented on Figures $1$ and $2$. It is clear that Eq.(43) does not have
positive constant solutions --- the width of the domain wall is
$\tau$-dependent.

Moreover, one can also develop an analytic, perturbative method for
investigating solutions of Eq.(43). It is based on the following
observation. For $\kappa\rightarrow\infty$ Eq.(43) possesses a constant
solution $A=1$. Because of the relationship of the function $A$ with
the width of the domain wall, formula (40), the other constant solutions
in this limit, i.e. $A=0,\;A=-1,$ are not interesting. $A=1$ means that the
half-width of the domain wall is equal to $\alpha^{-1}$. For $\kappa$ large
but finite consider Eq.(43) in the linear approximation. Then $A=1+\delta A$,
where $\delta A$ is small, and Eq.(43) is reduced to
\[
\frac{1}{2\kappa^{2} \cos^{2}{t}}\frac{d^2 \delta A}{dt^2} = -\frac{1}
{\kappa^{2}
\cos^{4}{t}}
-\frac{\delta A}{\kappa^{2}\cos^{4}{t}} - 2\delta A.
\]
This equation has the form of equation of motion for a linear oscillator
with the external force $-(\kappa^{2}\cos^{4}{t})^{-1}$ and with time
dependent coefficients. However this time dependence should be regarded
as very slow (adiabatic), because we are considering the variable $t$
only in the
interval (44) so the cosinus does not oscillate, and $\kappa$ is assumed to
be so large that $\delta A$ oscillates many times in that interval. For
example,
for $\kappa \simeq 30$ there are about ten full cycles of oscillations
of $\delta A$ for $t$ in the interval [0,1] (see Figure $3$). Also
the external force is rather
small for large $\kappa$ --- we expect that its presence results in a slow
drift of average value of $\delta A$ towards negative values. Therefore it
seems
natural to split $\delta A(t)$ into two components: the oscillating one with
large frequency of the order $2\kappa$, and the non-oscillating one of the
order of the external force i.e. $\kappa^{-2}$.

 This idea of splitting the inverse, dimensionless half-width
of the domain wall $A(t)$ into two components
is crucial for formulation of the perturbative  method of computing it.
We write
\begin{equation}
A(t)= N(t) + \Omega(t),
\end{equation}
where $N(t)$ is the non-oscillating component while $\Omega(t)$ is the
oscillating one, and we split Eq.(43) into two equations
\begin{equation}
\frac{1}{2\kappa^{2}\cos^{2}{t}}\frac{d^{2}N}{dt^{2}}=-\frac{N}{\kappa^{2}
\cos^{4}{t}} +N-N^{3},
\end{equation}
\begin{equation}
\frac{1}{4\kappa^{2}\cos^{2}{t}}\frac{d^{2}\Omega}{dt^{2}}=-\frac{1}{2}\left(
\frac{1}{\kappa^{2}\cos^{4}{t}}-1+\Omega^{2}+3\Omega N +3N^{2}\right)\Omega.
\end{equation}
If $N$ and $\Omega$ obey Eqs.(46),(47) then $A=N+\Omega$ obeys Eq.(43).
The advantage of replacing Eq.(43) by the two Eqs.(46),(47) is that each
of the two equations can be solved with the use of different perturbative
schemes. Let us assume for a moment that $N(t)$ and $\Omega(t)$ are
generic functions (of the indicated types) of the order $1$.
Then, for the non-oscillating component the expression
$(2\kappa^{2}\cos^{2}{t})^{-1}d^2N/dt^2$ present on the l.h.s. of Eq.(46) is
of the order $\kappa^{-2}$, while $(4\kappa^{2}\cos^{2}{t})^{-1}
d^2 \Omega/dt^2$
for the component oscillating with the frequency $\sim 2\kappa$ is of the
order $\kappa^{0}$ because $d^2\Omega/dt^2\sim 4\kappa^{2}$.

Therefore, it is more appropriate to write Eq.(46) in the form
\begin{equation}
N-N^{3}= \frac{1}{\kappa^{2}}\left(\frac{N}{\cos^{4}{t}}+\frac{1}
{2 \cos^{2}{t}}\frac{d^2 N}{dt^2}\right),
\end{equation}
and to seek $N$ in the form
\begin{equation}
N(t)=1+\kappa^{-2}N^{(2)}(t)+\kappa^{-4}N^{(4)}(t) + \ldots,
\end{equation}
regarding the whole r.h.s. of Eq.(48) as the perturbation. This gives
\begin{equation}
N^{(2)}(t)=-\frac{1}{2\cos^{4}{t}}, \;\; N^{(4)}(t)=-\frac{2}{\cos^{6}{t}}+
\frac{19}{8 \cos^{8}{t}}.
\end{equation}

On the other hand, in order to calculate $\Omega(t)$ we assume that
\begin{equation}
\Omega(t) = \kappa^{-2} \Omega^{(2)}(t) +\kappa^{-4}\Omega^{(4)}(t)+ \ldots,
\end{equation}
and we solve Eq.(47) order by order in $\kappa^{-2}$ regarding $\kappa^{-2}
\ddot{\Omega}^{(2k)},\:k=1,2,\ldots $ as of the order $\kappa^{-2k}$. For
$N(t)$ present on the r.h.s. of Eq.(47) we use the expansion (49). In this
way we obtain the following equations for $\Omega^{(2)}$ and $\Omega^{(4)}$:
\begin{equation}
\frac{1}{4\kappa^{2}\cos^{2}{t}}\frac{d^2 \Omega^{(2)}}{dt^2} =
-\Omega^{(2)},
\end{equation}
\begin{equation}
\frac{1}{4\kappa^{2}\cos^{2}{t}}\frac{d^2\Omega^{(4)}}{dt^2}=-\Omega^{(4)}
- \left(\frac{3}{2}\Omega^{(2)}-\frac{1}{\cos^{4}{t}}\right)\Omega^{(2)} .
\end{equation}
In the last equation we have used the result (50) for $N^{(2)}$.
Equation (52) is actually an example of the Mathieu-equation \cite{Abra}.
For our purposes, however, we will only need the following approximate solution
\begin{equation}
\Omega^{(2)}(t)=\frac{1}{\cos{t}}\left[c_{1}\sin{(2\kappa\sin{t})}+ c_{2}
\cos{(2\kappa\sin{t})}\right],
\end{equation}
where $c_{1},\:c_{2}$ are constants to be determined from initial data for
$A(t)$. The function $\Omega^{(2)}(t)$ given by formula (54) obeys Eq.(52)
up to terms of the order $\kappa^{-2}$, so it is sufficiently accurate for
solution in the order $\kappa^{-2}$. Of course we assume that the constants
$c_{1},\:c_{2}$ are of the order $1(=(\kappa^{-2})^{0})$ at most. Otherwise
$\Omega^{(2)}$ would be too large as for the first order correction.

The function $A(t)$ up to the order $\kappa^{-2}$ is given by
\begin{eqnarray}
A(t)&=&1+N^{(2)}(t)+\Omega^{(2)}(t)  \\
   & =&1- \frac{1}{2\kappa^{2}\cos^{4}{t}} + \frac{1}{\kappa^{2}\cos{t}}\left[
    c_{1}\sin{(2\kappa\sin{t})} +c_{2}\cos{(2\kappa\sin{t})}\right]. \nonumber
 \end{eqnarray}
It follows from formula (55) that the constants $c_{1},\:c_{2}$ have the
following relations to the initial data for $A(t)$:
\begin{equation}
A(0)=1+\frac{1}{\kappa^{2}}\left(c_{2}-\frac{1}{2}\right),\;\;
\frac{dA}{dt}(0) = \frac{2c_{1}}{\kappa}.
\end{equation}
Because $c_{1},\:c_{2}$ are typically of the order 1, $A(0)$ differs
from 1 by a number of the order $\kappa^{-2}$, while $dA/dt(0)$ is of the
order $\kappa^{-1}$. Formula (55) agrees rather well with the numerical
solutions of Eq.(43), see Figure 4 and compare with Figure 1.

Notice that we can solve Eq.(47) by putting $\Omega(t)=0$. Then $A(t)=N(t)$,
where $N(t)$ is given by formulae (49),(50). This solution contains
no arbitrary constant, and there is no freedom in the choice of
the initial data. To the order $\kappa^{-4}$ the initial data are
\begin{equation}
A(0)=1-\frac{1}{2\kappa^{2}}+\frac{3}{8\kappa^{4}},\;\;
\frac{dA}{dt}(0)=0.
\end{equation}
This very particular solution with no oscillatory component can  be
obtained by perturbative expansion in the positive powers of $\kappa^{-2}$ in
the original equation (43).
It is a very interesting fact that such solution  exists. On the other hand,
 it is just a single, isolated solution --- a generic cylindrical domain
 wall solution contains the oscillatory component.

 The presence of the oscillatory mode does not have any influence on
 the motion of the core, which in all cases is described by Eq.(29).

  As for any perturbative solution, also in our case there is the problem
  of convergence of the perturbative series. We shall not attempt to
  approach this problem from the rigorous, mathematical standpoint.
  We shall be satisfied with the observation that there is a hope
  for convergence of the series if $t$ is not too large (within the
  interval (44)). We see from  (49), (50) that for $N(t)$ the
   actual expansion parameter is $(\kappa^{2}\cos^{4}{t})^{-1}$.
 For $\Omega(t)$, (51), (52) suggest that the actual expansion
 parameter is $(\kappa^{2}\cos{t})^{-1}$. Because $t$ changes from
 0 to $\frac{\pi}{2}-\delta$, see (44), the former expansion parameter
 is bigger than the latter one. Therefore we expect that the difference
 between $A(t)$ in the order $\kappa^{-2}$ (given by formula (55)), and the
exact solution of Eq.(43), is of the order $(\kappa^{2}\cos^{4}{t})^{-2}$.
This number increases with $t$ and thus it is maximal for  $t=\frac{\pi}{2}
-\delta$. Taking $\delta$ determined from formula (59) below, we find that
the maximal value of that number is equal to  $d^{-4}$, which is
approximately 0.35.

{}From formulae (40) and (55) we obtain the half-width of the domain wall
in the order $\kappa^{-2}$:
\begin{equation}
\xi_{0}(t)= \alpha^{-1}\left[ 1+\frac{1}{2\kappa^{2}\cos^{4}{t}} -
\frac{1}{\kappa^{2}\cos{t}}(c_{1}\sin{(2\kappa\sin{t})} +c_{2}\cos{(2\kappa
\sin{t})})\right].
\end{equation}
Its initial value is
\[ \xi_{0}(0)= \alpha^{-1}(1+\frac{1}{2\kappa^{2}} -
\frac{c_{2}}{\kappa^{2}}).  \]
If $c_{1},\; c_{2}$ are not too big, $\xi_{0}(t)$ increases with $t$.

Because we now know $\xi_{0}(\tau)$, we can estimate the value of $\delta$
present in formula (38). Formulae (37), (38)  give
\[ \kappa \cos^{2}{(\frac{\pi}{2}-\delta)}\approx 1+\frac{1}{2\kappa^{2}
\cos^{4}{(\frac{\pi}{2}-\delta)}} - \frac{c_{1}\sin{(2\kappa\sin{t})}+c_{2}
\cos{(2\kappa\sin{t})}}{\kappa^{2}\cos{(\frac{\pi}{2}-\delta)}}.  \]
This gives
\begin{equation}
\sin^{2}{\delta}\approx \frac{d}{\kappa},
\end{equation}
where $d\approx 1.297$ is determined from the equation
\[ d=1+\frac{1}{2d^{2}}. \]
For example, if $\kappa=3$ then $\delta\approx 0.72$, if $\kappa=30$ then
$\delta \approx 0.21$. For $t\geq t_{\mbox{max}}\equiv\frac{\pi}{2}-\delta $
our solutions
loose their meaning as describing the evolution of the domain wall. For
$\kappa =3 \;\; t_{\mbox{max}}\approx 0.85$, and for $\kappa =30 \;\;
 t_{\mbox{max}}\approx 1.36$. For comparison, if the width of the domain wall
 was constant in $\tau$, e.g. equal to $\alpha^{-1}$, then the
 condition (37) would give an equation for $\delta$ of the form (59) but with
 $d=1$. In this case, for $\kappa=3$ $t_{\mbox{max}}\approx 0.96$, and for
 $\kappa=30$ $t_{\mbox{max}}\approx 1.39$. Thus, the effect of broadening
 of the wall is more significant for small $\kappa$.

\section{Transformation to the laboratory frame }

The relation of the coordinates $(\tau,\sigma,\xi,\zeta)$ to the laboratory
frame coordinates is given by formula (7). It is non-trivial, in particular
 it requires the knowledge of $\tau$-evolution of the core of the
 domain wall. This Section is devoted to a detailed analysis of this
 intricate relation.

  The field $\Phi$ is a scalar field, so in order to transform it it is
  sufficient to express the $\tau,\xi$ variables by the laboratory frame
  coordinates $x^{\mu}$:
  \[ \Phi(\tau,\xi) \rightarrow \tilde{\Phi}(x^{0},x^{1},x^{2},x^{3})=
  \Phi(\tau(x^{\mu}),\xi(x^{\mu})).    \]

   With the core of the domain wall
contracting according to formulae (34),(35), formula (7) gives
the lab-frame coordinates as explicit functions of the special coordinates:
\begin{eqnarray}
x^{0}=\tau - \xi \tan{\frac{\tau}{r_{0}}} , \\
\rho\equiv \sqrt{(x^{1})^{2}+(x^{2})^{2}} = r_{0}\cos{\frac{\tau}{r_{0}}}
+ \frac{\xi}{\cos{\frac{\tau}{r_{0}}}},  \\
\theta \equiv \arctan{\frac{x^{2}}{x^{1}}}=\sigma, \;\;\; x^{3}=\xi,  \nonumber
\end{eqnarray}
where again we have put $\tau_{0}=0$. Here we have passed to the polar
coordinates $(\rho,\theta)$ in the $(x^{1},x^{2})$ plane.
Relations (60),(61) are simple as far as
dependence on $\xi$ is concerned; on the other hand dependence on  $\tau$
is quite nontrivial. Only for $\xi=0$, i.e. on the core of the domain wall,
the parameter $\tau$ coincides with the laboratory time $x^{0}$. In general,
 simultaneity in $\tau$ is not equivalent to simultaneity in the lab-frame
 time $x^{0}$.

 The interval of $\tau$ we are interested in is given by (38), with the
 estimate for $\delta$ given by (59). The allowed range of $\xi$ is
  $\tau$-dependent:
  \begin{equation}
  \mid\xi\mid < r_{0}\cos^{2}{\frac{\tau}{r_{0}}},
  \end{equation}
  as follows from  (11) in the case of solution (34) for the motion of the
  core. We assume that outside of this range of $\xi$ the $\Phi$ field takes
  one of the vacuum values. Notice that the range of $\xi$ shrinks faster
  than the core of the wall, and it would shrink to a point if $\tau$ could
  approach $\pi r_{0}/2$. The interval $0\leq\tau<\frac{\pi}{2}r_{0}$
  corresponds to $0\leq x^{0}<\frac{\pi}{2}r_{0}$. The interval (62) of the
  variable $\xi$ corresponds to $0<\rho<2r_{0}\cos{\frac{\tau}{r_{0}}}$, so
  it shrinks too. For any $\xi$ in the interval (62) $dx_{0}/d\tau$ is
  positive, so the relation between $\tau$ and $x_{0}$ is one-to-one.

  Let us analyse the relations (60),(61) in a vicinity of the core, where
  $\xi/r_{0}$ is small. For such $\xi$ we can give approximate analytic
  formula for $\tau$ as a function of $x^{0}$ and $\rho$. To this end we
  write the formula (60) in the following form
  \begin{equation}
\frac{\tau}{r_{0}}=\frac{x^{0}}{r_{0}}+\frac{\xi}{r_{0}}
\tan{\frac{\tau}{r_{0}}},
\end{equation}
and we solve it iteratively regarding the term $(\xi/r_{0})\tan{\tau/r_{0}}$
as a perturbation. This is justified because due to the restriction (62) on
the range of $\xi$ that term is bounded by $\frac{1}{2}\mid\sin{\frac{2\tau}
{r_{0}}}\mid $ for any $\tau$, and for $\xi$ sufficiently close to 0 it
becomes a small perturbation. Denoting by $\tau^{(i)}$ the result of i-th
iteration we find
\begin{eqnarray}
\tau^{(3)}&=& x^{0}+\xi\tan{\frac{x^{0}}{r_{0}}}\left[
1+\frac{\xi}{r_{0}\cos^{2}{\frac{x^{0}}{r_{0}}}}   \right.    \\
 & &\left. \mbox{} +\frac{\xi^{2}}{r_{0}^{2}\cos^{4}{\frac{x^{0}}{r_{0}}}}
 \left(2-\cos^{2}{\frac{x^{0}}{r_{0}}}\right)
 \right] + {\cal O}(\frac{\xi^{4}}{r_{0}^{3}}).
\nonumber
\end{eqnarray}
Using formulae (61),(64) we can calculate $\rho$ as a function of $x^{0}$
and $\xi$ up to the order $\xi^{3}$:
\begin{eqnarray}
\rho&=& \cos{\frac{x^{0}}{r_{0}}} \left[ r_{0}+\xi
-\frac{\xi^{2}}{2r_{0}}\tan^{2}{\frac{x^{0}}{r_{0}}} \right.   \\
& & \left. \mbox{} - \frac{\xi^{3}}{6r_{0}^{2}}\tan^{2}{\frac{x^{0}}{r_{0}}}
\left(1+\frac{2}{\cos^{2}{\frac{x^{0}}{r_{0}}}}\right) \right] +{\cal
O}(\frac{\xi^{4}}{r_{0}^{3}}).
\nonumber
\end{eqnarray}
This formula can easily be inverted, yielding  $\xi$ as a function of $\rho$
and $x^{0}$: with the notation
\[ \overline{\rho}\equiv \rho - r_{0}\cos{\frac{x^{0}}{r_{0}}} \]
we have
\begin{eqnarray}
\xi&=&\frac{\overline{\rho}}{\cos{\frac{x^{0}}{r_{0}}}}  \left[  1 +
\frac{(\overline{\rho})}
{2r_{0}\cos{\frac{x^{0}}{r_{0}}}}
\tan^{2}{\frac{x^{0}}{r_{0}}} \right.   \\
& & \left. \mbox{}+\frac{(\overline{\rho})^{2}}{3r_{0}^{2}
\cos^{2}{\frac{x^{0}}{r_{0}}}}
\left(\frac{5}{2\cos^{2}{\frac{x^{0}}{r_{0}}}}-1\right)
\tan^{2}{\frac{x^{0}}{r_{0}}}  \right]
+ {\cal O}((\overline{\rho})^{4}). \nonumber
\end{eqnarray}
Inserting this expression on the r.h.s. of formula (64) we obtain $\tau^{(3)}$
as the function of the lab-frame coordinates
\begin{eqnarray}
\tau^{(3)}&=& x^{0}+\overline{\rho} \frac{1}{\cos{\frac{x^{0}}{r_{0}}}}
\tan{\frac{x^{0}}{r_{0}}} \left[ 1+\frac{(\overline{\rho})}{2r_{0}\cos^{3}
{\frac{x^{0}}{r_{0}}}} \left( 3-\cos^{2}{\frac{x^{0}}{r_{0}}}\right)
\right. \\
& &\left. \mbox{}+\frac{(\overline{\rho})^{2}}{3r_{0}^{2}\cos^{6}
{\frac{x^{0}}{r_{0}}}} \left(\frac{23}{2}-\frac{19}{2}
\cos^{2}{\frac{x^{0}}{r_{0}}}+\cos^{4}{\frac{x^{0}}{r_{0}}}\right) \right]
+{\cal O}((\overline{\rho})^{4}).  \nonumber
\end{eqnarray}

It is interesting to have a look at the laboratory frame radial velocity
of a point of the wall with given fixed $\xi$:
\begin{eqnarray}
\frac{d\rho}{dx^{0}}\mid_{\xi=\mbox{const}}& = & -\sin{\frac{x^{0}}{r_{0}}}
\left[ 1+\frac{\xi}{r_{0}}+\frac{\xi^{2}}{2r_{0}^{2}}\left(1+\frac{1}
{\cos^{2}{\frac{x^{0}}{r_{0}}}}\right) \right.  \\
& &\left.\mbox{}+\frac{\xi^{3}}{6r_{0}^{3}}\left(  1
-\frac{1}{\cos^{2}{\frac{x^{0}}{r_{0}}}}
 + \frac{6}{\cos^{4}{\frac{x^{0}}{r_{0}}}}   \right)
\right] + {\cal O}((\frac{\xi}{r_{0}})^{4}). \nonumber
\end{eqnarray}
This formula for $\xi=0$ gives the lab-frame velocity of the core of the
domain wall. We see that points lying inside the cylinder formed by the core
($\xi<0$) move radially inward with smaller velocity than points from the
outside of the cylinder ($\xi>0$). The r.h.s. of formula (68) is equal
 to $-\sin{\frac{\tau}{r_{0}}}$, with $\tau=\tau^{(3)}$ given by formula
 (64). Actually one can derive directly from formulae (60),(61) that
 \begin{equation}
 \frac{d\rho}{dx^{0}}\mid_{\xi=\mbox{const}}=-\sin{\frac{\tau}{r_{0}}}.
 \end{equation}

 From formulae (60),(61) also follows the relation
 \begin{equation}
 d\rho\mid_{x^{0}=\mbox{const}}=\cos{\frac{\tau}{r_{0}}}\;d\xi
 \end{equation}
 (in the derivation of it it is crucial to observe that $d\xi\neq 0$ implies
 $d\tau\neq 0$ because $x^{0}$ is kept constant). In view of formula (69),
  this means that $d\rho$ in the lab-frame and $d\xi$ in the special are
  related by the local boost with velocity
$d\rho/dx^{0}\mid_{\xi=\mbox{const}}$,
  because $\cos{\frac{\tau}{r_{0}}}$ is equal to the corresponding Lorentz
  contraction factor.

  In order to obtain the width $\Delta \rho$ of the domain wall in the
  laboratory frame we integrate formula (70) in the interval
  $[-\xi_{0},\xi_{0}]$, with $\tau$ given by formula (64). This yields
  \begin{equation}
  \Delta\rho = 2\xi_{0} \cos{\frac{x^{0}}{r_{0}}}\left[
  1-\frac{\xi^{2}_{0}}{3r^{2}_{0}}
  \tan^{2}{\frac{x^{0}}{r_{0}}} \left(  \frac{1}{2}+\frac{1}
  {\cos^{2}{\frac{x^{0}}{r_{0}}}}\right)\right].
  \end{equation}
  The same result is obtained from formula
  \begin{equation}
  \Delta\rho=\rho(\xi_{0},x^{0})-\rho(-\xi_{0},x^{0}),
  \end{equation}
  where $\rho(\xi,x^{0})$ is given by formula (65).

  The factor $\cos{(x^{0}/r_{0})}$ in formula (71) can be regarded as an
  "average" Lorentz contraction factor -- it is exactly the Lorentz contraction
  factor for the radius of the core of the wall. The
  term $2\xi_{0}$ gives the width of the
  wall in the special coordinate system. The correction term (proportional to
  $\xi^{3}_{0}$) is due to the difference in the lab-frame velocities of the
  points of the domain wall. It is easy to see that if $\xi_{0}$ was
  constant in time, the presence of this term would result in decreasing  the
  lab-frame width $\Delta\rho$ (apart from the Lorentz contraction). However,
   we have found that $\xi_{0}$ depends on $\tau$: formulae (40),(55) give
   \begin{equation}
   \alpha\xi_{0}(\tau)=1+\frac{1}{2\kappa^{2}\cos^{4}{\frac{\tau}{r_{0}}}}
   -\frac{c_{1}\sin{(2\kappa\sin{\frac{\tau}{r_{0}}})}+c_{2}

\cos{(2\kappa\sin{\frac{\tau}{r_{0}}})}}{\kappa^{2}\cos{\frac{\tau}{r_{0}}}}.
   \end{equation}
   In order to pass to the laboratory frame coordinates, we have to
   eliminate $\tau$ in favour of $x^{0}$ in this formula. To this end, we
   use formula (64), in which we put $\xi=\xi_{0}(\tau)$, because
   $\xi_{0}(\tau)$ is the value of the coordinate $\xi$ at which
   $\Phi(\xi,\tau)=+v$. Then formula (73) becomes a complicated equation
   from which one should determine $\xi_{0}$ as function of $x^{0}$.
   This function we shall denote by $\tilde{\xi}_{0}(x^{0})$ --- thus
   $ \tilde{\xi}_{0}(x^{0})\equiv \xi_{0}(\tau)$.

   Actually, this task is quite simple if we recall that formula (73)
    is obtained in  the perturbative expansion up to the order
     $\kappa^{-2}$ -- terms of the order $\kappa^{-4}$ are not
     included. From formula (64) we obtain (for $\xi=\xi_{0}(\tau)$)
   \begin{eqnarray}
   \cos{\frac{\tau^{(3)}}{r_{0}}} &=&\cos{\frac{x^{0}}{r_{0}}}\left[ 1-
   \frac{\alpha\xi_{0}(\tau)}{\kappa}\tan^{2}{\frac{x^{0}}{r_{0}}}
   -\frac{(\alpha\xi_{0})^{2}}{\kappa^{2}}\left(\frac{1}{2}+\frac{1}{\cos^{2}
   {\frac{x^{0}}{r_{0}}}}\right)\tan^{2}{\frac{x^{0}}{r_{0}}} \right.
\nonumber        \\
& & \mbox{} \left. -\frac{(\alpha\xi_{0})^{3}}{\kappa^{3}}(...)\right] +{\cal
O}
   (\frac{(\alpha\xi_{0})^{4}}{\kappa^{4}}).
      \end{eqnarray}
   The corresponding formula for $\sin{(\tau/r_{0})}$ is
   obtained from (74) and $\sin{x}= \\ \sqrt{1-\cos^{2}{x}}$. It is clear that
   the terms in formula (74) which contain $\kappa^{-1}, \;\kappa^{-2},$ etc,
   will give contributions to the r.h.s. of formula (73) of the order
   higher than $\kappa^{-2}$. Therefore, in order to obtain
   $\alpha\tilde{\xi}_{0}(x^{0})$  to the order $\kappa^{-2}$ we may take
   \begin{equation}
   \cos{\frac{\tau^{(3)}}{r_{0}}}\approx  \cos{\frac{x^{0}}{r_{0}}}.
   \end{equation}
   Consequently
   \begin{equation}

\alpha\tilde{\xi}_{0}(x^{0})=1+\frac{1}{2\kappa^{2}\cos^{4}{\frac{x^{0}}{r_{0}}}}
   -\frac{c_{1}\sin{(2\kappa\sin{\frac{x^{0}}{r_{0}}})}+
   c_{2}\cos{(2\kappa\sin{\frac{x^{0}}{r_{0}}})}}
   {\kappa^{2}\cos{\frac{x^{0}}{r_{0}}}}    +{\cal O}(\kappa^{-3}).
   \end{equation}
   Inserting this in formula (71) we obtain the time dependence
   of the laboratory frame width of the domain wall:
   \begin{eqnarray}
   \alpha\Delta\rho &=&\cos{\frac{x^{0}}{r_{0}}}\left[2+\frac{1}{3\kappa^{2}}
   \left(1+
   \frac{1}{\cos^{2}{\frac{x^{0}}{r_{0}}}}+\frac{1}{\cos^{4}
   {\frac{x^{0}}{r_{0}}}}\right) \right.   \\
    & & \mbox{}\left. -\frac{2}{\kappa^{2}}\frac{c_{1}\sin{(2\kappa
   \sin{\frac{x^{0}}{r^{0}}})}+c_{2}\cos{(2\kappa\sin{\frac{x^{0}}{r_{0}}})}}
   {\cos{\frac{x^{0}}{r_{0}}}}\right] +{\cal O}(\kappa^{-3}).   \nonumber
   \end{eqnarray}
   It is easy to see that if $c_{1},\;c_{2}$ are not too large,
   the expression in the square bracket increases with $x^{0}$, in
   contradistinction to the earlier mentioned case of $\xi_{0}$ constant
   in $\tau$. Of course, due to the Lorentz contraction factor
   $\cos{\frac{x^{0}}{r_{0}}}$, the laboratory frame width of the domain wall
   decreases also in the present case.

   Finally, let us present the form of the field $\tilde{\Phi}(x^{0},\rho,
   \theta,x^{3})$ of the domain wall in the laboratory frame:
   \begin{eqnarray}
   \tilde{\Phi}= \left\{  \begin{array}{ccl}
   +v & \mbox{for} & \rho\geq \rho_{(+)}, \\
   -v & \mbox{for} & \rho\leq\rho_{(-)}.
   \end{array}
   \right.
   \end{eqnarray}
Here $\rho_{(+)},\; \rho_{(-)} $ are the outer and the inner
lab-frame radia of the cylindrical domain wall,
respectively. They can be calculated from the relations
\begin{equation}
\alpha\xi(x^{0},\rho_{(+)}) = \alpha\tilde{\xi}_{0}(x^{0}), \;\;\;
\alpha\xi(x^{0},\rho_{(-)})=-\alpha\tilde{\xi}_{0}(x^{0}),
\end{equation}
where $\xi(x^{0},\rho),\;\tilde{\xi}_{0}(x^{0})$ are given by (66),(76),
respectively. Solving (79) perturbatively in $\kappa^{-1}$ we obtain the
following formulae (to the order $\kappa^{-2}$)
\begin{eqnarray}
\alpha\rho_{(\pm)} &=& \cos{\frac{x^{0}}{r_{0}}}\left[\alpha r_{0}\pm 1 -
\frac{1}{2\kappa}\tan^{2}{\frac{x^{0}}{r_{0}}}
 \pm \frac{1}{6\kappa^{2}}\left(\cos^{-4}{\frac{x^{0}}{r_{0}}}
  +\cos^{-2}{\frac{x^{0}}{r_{0}}}+1\right) \right.  \nonumber      \\
 & &\mbox{}\left. \mp \frac{c_{1}\sin{(2\kappa\sin{\frac{x^{0}}{r_{0}})}}+
  c_{2}\cos{(2\kappa\sin{\frac{x^{0}}{r_{0}}})}}
  {\kappa^{2}\cos{\frac{x^{0}}{r_{0}}}} \right].
\end{eqnarray}
For intermediate values of $\rho$, i.e. $\rho\in(\rho_{(-)},\rho_{(+)})$,
\begin{equation}
\tilde{\Phi}= \frac{3}{2} v \frac{\xi(x^{0},\rho)}{\tilde{\xi}_{0}(x^{0})}
\left( 1-\frac{1}{3}\frac{\xi^{2}(x^{0},\rho)}{\tilde{\xi}^{2}_{0}(x^{0})}
 \right),
 \end{equation}
 as follows from formulae (19),(28) and (31). Here for
 $\xi(x^{0},\rho)$ and $\tilde{\xi}_{0}(x^{0})$ one should substitute
 the expressions (66) and (76).

\section{Remarks}

 The main new result about the  physics of the domain walls
  obtained in our approach is the observation
that in general there is the oscillatory component $\Omega(\tau)$.
The presence of
this component manifests itself not only as the oscillations of the width
of the domain wall --- from formulae (19),(28) we see that for each fixed
$\xi$ in the domain wall (i.e. $\xi\in(-\xi_{0},\xi_{0})$) the value
of the field $\Phi(\tau,\xi)$ oscillates too. When the oscillatory
mode is taken into account, the field $\Phi$ becomes a singular
function of the parameter $\kappa^{-1}$ at $\kappa\rightarrow \infty$,
 so it can not be obtained in a perturbative expansion in $\kappa^{-1}$.

 As merits of our approach we would like to mention its computational
simplicity,
  the fact that the width of the domain wall appears explicitly as
  a dynamical variable, and the fact that we can compute the oscillatory
  component. Of course there are shortcomings too. The most troublesome one
  is the lack of a physical "control parameter" in the approximation
  consisting in cutting the Taylor series. There is no physical
  parameter, let us denote it by $\chi$, such that for,e.g. $\chi
  \rightarrow 0$ our approximate solution (81) becomes exact, and that
  deviation from the exact solution for finite $\chi$ is of the order,
  e.g. $\chi^{n}$ with some $n$. (Notice however, that for Eq.(39)
  for the inverse width of the domain wall we have found the relevant
  physical control parameter -- it is $\kappa^{-2}$.)
   Therefore, in our approach it is much
  harder than usually to estimate the accuracy of our solution, and
  to identify the physically relevant parts in expressions for energy
  and momentum of the wall obtained from the approximate solution. We
  would like to describe this problem in more detail using the energy as
  the example, but before that let us digress on how to define a
  convenient energy-momentum four-vector.

 One possibility is to use directly
 the energy-momentum  tensor $T^{\mu\nu}$ given by (4) and to compute the
 energy in the laboratory frame. However, this leads to rather tedious
 computations because the transformation to the laboratory frame is not
 simple, as we have seen in Section 4. More convenient approach is to
 pass to quantities which are constant with respect to the $\tau$ parameter.
  These quantities are obtained by integration of appropriately projected
  components of the $T^{\mu\nu}$ tensor over the hypersurfaces
  $\tau=\mbox{const.}$ More specifically, the continuity equation
  \begin{equation}
  \partial_{\mu}T^{\mu\nu}=0
  \end{equation}
  is equivalent to
  \begin{equation}
    \partial_{\alpha}(\sqrt{-G}T^{\alpha\nu})=0,
      \end{equation}
 where $\partial_{\alpha}\equiv \partial/\partial u^{\alpha}, (u^{\alpha})=
(\tau,\sigma,\zeta,\xi)$ is another notation for the special coordinates,
and
\begin{equation}
T^{\alpha \nu}= \frac{\partial u^{\alpha}}{\partial x^{\mu}} T^{\mu\nu}.
\end{equation}

Notice that in (84) we have transformed to the special coordinates only one
index of $T^{\mu\nu}$. The point is that then Eq.(83) has the form of
vanishing of the ordinary divergence. With the two indices of $T^{\mu\nu}$
transformed, the continuity equation (82) can not be written in the form of
vanishing of the ordinary divergence --- the covariant divergence of the
second rank tensor can not be written in the form of the ordinary divergence,
in general.

Because in the case of cylindrical domain wall the field $\Phi$ depends
only on $\tau$ and $\xi$, and $\partial u^{\alpha}/\partial x^{\mu}$ depends
on $\tau,\xi$ and $\sigma$, equation (83) reduces to
\begin{equation}
\partial_{\tau}(\sqrt{-G}T^{\tau\nu}) +
\partial_{\sigma}(\sqrt{-G}T^{\sigma\nu}) + \partial_{\xi}(\sqrt{-G}
T^{\xi\nu}) =0.
\end{equation}
It is also easy to see that for our solutions $T^{\mu\nu}$ vanishes for
$|\xi|>|\xi_{0}|$. Therefore, the quantities
\begin{equation}
P^{\nu}\equiv -\int^{\xi_{0}}_{-\xi_{0}}\;d\xi \int^{2\pi}_{0}\;d\sigma
\sqrt{-G} T^{\tau\nu},
\end{equation}
(where $\nu=0,1,2,3$) are constant in the $\tau$ parameter. $P^{0}$ can be
regarded as the energy per unit length of the cylinder. Its density
is proportional to the projection of $T^{\mu 0}$ on the four-vectors
$\partial\tau/\partial x^{\mu}$, which are orthogonal to the hypersurface
$\tau=\mbox{const}$. Simple computation shows that
\begin{eqnarray}
\lefteqn{-\sqrt{-G}T^{\tau 0}= } \nonumber \\
 & &\frac{r}{\sqrt{1-\dot{r}^{2}}}
  \left(1+\frac{\xi}{r\sqrt{1-\dot{r}^{2}}}\right) \left\{
  \frac{1}{2}\frac{1}{1-\dot{r}^{2}}
\left(1+\frac{\xi \ddot{r}}{(\sqrt{1-\dot{r}^{2}})^{3}}
\right)^{-2}\Phi_{,\tau}^{2} \right.\nonumber \\
 & & \left. \mbox{} -\frac{\dot{r}}{\sqrt{1-\dot{r}^{2}}}
  \left(1+\frac{\xi\ddot{r}}{(\sqrt{1-\dot{r}^{2}})^{3}}\right)^{-1}
 \Phi_{,\tau} \Phi_{,\xi}   + V(\Phi)+\frac{1}{2}
(\partial_{\xi}\Phi)^{2}  \right\} .
\end{eqnarray}
Notice that the  factor in front of the  r.h.s. of
this formula is equal to the Nambu-Goto energy (32).

Now, we would like to substitute for $\Phi(\tau,\xi)$ our
approximate solution (19),(28), (58), for $r(\tau)$ the
Nambu-Goto trajectory (34), and to compute $P^{0}$.
Because our solution is only the approximate one, the computed
$P^{0}_{(\mbox{appr})}$ is not constant in $\tau$, because the continuity
equation (82) holds only for exact solutions of the field equation (3).
The question is what is the relation of the $P^{0}_{(\mbox{appr})}$ to
the energy $P^{0}$ for the exact solution to which our solution is an
approximation. Only after clarifying this point one can present the
physically meaningful approximate value of the energy, and, subsequently,
to answer more detailed questions, e.g. about  contribution of the
internal oscillations to the energy. We plan to address this issue in a
future paper.

There are also several other interesting topics to be addressed. One would
like to extend our calculations by including
further terms in the Taylor expansion (19). It might turn out that some
corrections to the Nambu-Goto equation (29) will appear.

Yet another task is to apply our Taylor expansion approach to generic, less
symmetrical domain walls in the model defined by Lagrangian (1),(2). Here
an interesting question is about presence of a dynamical coupling of
internal degrees of freedom to the core degrees of freedom. In the case
of symmetrical walls considered in the present paper such coupling is in fact
absent, eventhough the motion of the core gives a "background" for the
internal oscillations, see Eq.(39).

Finally we would like to apply our method to domain walls in other
field-theoretical models used in condensed matter physics.

\vskip 30pt
\hspace*{-6mm}{\bf Acknowledgements}
\vskip 6pt
\hspace*{-6mm}H.Arod\'{z} would like to thank the Niels Bohr Institute for kind
hospitality and support.
\newpage
\hspace*{-6mm}{\bf THE APPENDIX. The spherical domain wall}
\vskip 6pt
\hspace*{-6mm}In this appendix
we consider briefly the method described in Section 2, when
used on the spherically symmetric domain wall. Since everything proceeds
in the same way as for the cylindrical wall we just give the few main steps
of the calculation.

The spherical domain wall is parametrised by:
\begin{equation}
\left(
\begin{array}{c}
X^0\\
X^1\\
X^2\\
X^3
\end{array}
\right) (\tau, \sigma, \zeta) = \left(\begin{array}{c}\tau\\
r(\tau)\;\sin\sigma\;\cos\zeta\\r(\tau)\;\sin\sigma\;\sin\zeta\\
r(\tau)\;\cos\sigma\end{array}\right),
\end{equation}
where now: $0\leq\sigma\leq\pi$ and $0\leq\zeta<2\pi$. In new coordinates
$(\tau, \sigma, \zeta, \xi)$ we get the following non-vanishing
components of the metric tensor (cf. Eqs.(7),(8))
\begin{equation}
G_{\tau\tau}=-(\sqrt{1-\dot{r}^2}+\frac{\xi\ddot{r}}{1-\dot{r}^2})^2,\;\;
G_{\sigma\sigma}=(r+\frac{\xi}{\sqrt{1-\dot{r}^2}})^2,
\end{equation}
and
\begin{equation}
G_{\zeta\zeta}=(r+\frac{\xi}{\sqrt{1-\dot{r}^2}})^2\sin^2\sigma,\;\;
G_{\xi\xi}=1,
\end{equation}
so that
\begin{equation}
\sqrt{-G}=(\sqrt{1-\dot{r}^2}+\frac{\xi\ddot{r}}{1-\dot{r}^2})
(r+\frac{\xi}{\sqrt{1-\dot{r}^2}})^2\sin\sigma.
\end{equation}
Now, following exactly the same steps as in section 2, we find in analogy
to Eqs. (22) and (23)
\begin{equation}
b=-a\frac{1}{\sqrt{1-\dot{r}^2}}(\frac{2}{r}+\frac{\ddot{r}}{1-\dot{r}^2}),
\end{equation}
\begin{eqnarray}
c=\frac{\ddot{a}}{1-\dot{r}^2}+\frac{\dot{a}\dot{r}}{1-\dot{r}^2}
(\frac{2}{r}+\frac{\ddot{r}}{1-\dot{r}^2})+\frac{a}{1-\dot{r}^2}
(\frac{2}{r}+\frac{\ddot{r}}{1-\dot{r}^2})^2\\
+\frac{a}{1-\dot{r}^2}(\frac{2}{r}+\frac{\ddot{r}^2}{(1-\dot{r}^2)^2})
-2\lambda v^2 a.\nonumber
\end{eqnarray}
Using patching conditions similar to (21) we finally get
\begin{equation}
\xi_0=\xi_1,\;\;a=\frac{3v}{2\xi_0},\;\;b=0,\;\;c=-\frac{3v}{\xi^3_0},
\end{equation}
as well as
\begin{equation}
\frac{\ddot{r}}{1-\dot{r}^2}+\frac{2}{r}=0,
\end{equation}
corresponding to the equation of motion for a spherical membrane obtained
from the Nambu-Goto action. Finally the time dependence of $a$ is determined
by
\begin{equation}
\frac{\ddot{a}}{1-\dot{r}^2}+\frac{6a}{r^2 (1-\dot{r}^2)}-
2\lambda v^2 a+\frac{8}{9v^2}a^3=0,
\end{equation}
which is just slightly different from equation (30) for the cylinder. In the
case of (95) the "energy" integral of motion is (compare with Eqs.(29),(31))
\begin{equation}
E=\frac{r^2}{\sqrt{1-\dot{r}^2}}.
\end{equation}
Then Eq.(95) can be written
\begin{equation}
\ddot{r}=\frac{-2r^3}{E^2},
\end{equation}
which is solved by a Jacobian elliptic function \cite{Abra}
\begin{equation}
r(\tau)=r_0 CN[\sqrt{\frac{2}{E}}(\tau-\tau_0)\mid\frac{1}{2}],
\end{equation}
where $r_0=r(\tau_0)=\sqrt{E}$. The "velocity" $\dot{r}(\tau)$ vanishes at
$\tau=\tau_0$ and the spherical wall collapses at $\tau=\tau_0+
\sqrt{\frac{E}{2}}K(\frac{1}{2})$, where $K(m)$ is the complete
elliptic integral of the first kind \cite{Abra}.

The analog of equation (43) becomes (taking again $\tau_0=0$)
\begin{equation}
\frac{1}{\tilde{\kappa}^2}\frac{1}{CN^4[\tilde{t}\mid\frac{1}{2}]}
\frac{d^2A}{d\tilde{t}^2}=\frac{-3}{\tilde{\kappa}^2}
\frac{A}{CN^6[\tilde{t}\mid\frac{1}{2}]}+A-A^3,
\end{equation}
where A is still given by Eq.(40) but
\begin{equation}
\tilde{t}=\sqrt{\frac{2}{E}}\tau=\frac{\sqrt{2}\tau}{r_0},\;\;\;
\tilde{\kappa}=\sqrt{E}\alpha=r_0\alpha.
\end{equation}
Numerically integrating Eq.(100) leads to curves similar to the curves
obtained in the cylindrical wall case (Figures 1,2). We can therefore expect
that the analysis of Sections 3 and 4 goes through for the spherical wall
also, and with qualitatively similar results.

\newpage
\begin{centerline}
{\bf Figure captions}
\end{centerline}
\vskip 12pt
\hspace*{-6mm}Fig.1. Numerical solution of equation (43) for
$\kappa=30$ (a), $\kappa=10$ (b), $\kappa=7$ (c)
and $\kappa=3$ (d). On a large scale the functions are almost
constant for (approximately) $\kappa>5$ and $t<1$. See also the comments
after Eq.(59).
\vskip 12pt
\hspace*{-6mm}Fig.2. The same solutions as showed on Figure 1, but magnified.
Now we explicitly see the oscillatory part representing the fine structure of
the solutions.
\vskip 12pt
\hspace*{-6mm}Fig.3. The fine structure of the solution for $\kappa=30$
obtained in two different ways: (a) represents $A(t)$ obtained by
integrating equation (43) while (b) represents $1+\delta A(t)$ obtained by
integrating the linearised equation for $\delta A(t)$ There is almost no
detectable difference
until $t$ approaches $1$, so that $A(t)\approx 1+\delta A(t)$ with
$\delta A(t)$ obeying the linearised equation is an extremely
good approximation for $\kappa=30$.
\vskip 12pt
\hspace*{-6mm}Fig.4. The function $A(t)$ given explicitly by formula (55) for
the same values of $\kappa$ as in Figures 1 and 2. There is quite good
agreement with the exact (numerical) solutions shown in Figure 1 for
$\kappa>5$ and $t<1$. This actually also holds for the fine
 structure (not shown here).

\end{document}